\documentclass[11pt]{article}
\usepackage{natbib}
\usepackage{times}
\usepackage{geometry} 
\usepackage[utf8]{inputenc}
\usepackage{enumitem,amssymb}
\usepackage{ragged2e}
\newlist{thematic}{itemize}{8}
\setlist[thematic]{label=$\square$}
\usepackage{pifont}

\usepackage{graphicx}

\begin{document}

% Astro 2020 web page: http://sites.nationalacademies.org/SSB/CurrentProjects/SSB_185159

% Relevant formatting requirements: 5 pages for text, tables, and figures (cover page and references don't count)

{\raggedright
\huge
Astro2020: 
Activity, Project, and Statement of the Profession Consideration White Paper \linebreak

Future Uses of the LSST Facility: Input from the LSST Project Science Team \linebreak

\normalsize
  
\noindent \textbf{Thematic Areas:} \hspace*{60pt} $\boxtimes$ Planetary Systems \hspace*{10pt} $\boxtimes$ Star and Planet Formation \hspace*{20pt}\linebreak
$\boxtimes$ Formation and Evolution of Compact Objects \hspace*{31pt} $\boxtimes$ Cosmology and Fundamental Physics \linebreak
  $\boxtimes$  Stars and Stellar Evolution \hspace*{1pt} $\boxtimes$ Resolved Stellar Populations and their Environments \hspace*{40pt} \linebreak
  $\boxtimes$    Galaxy Evolution   \hspace*{45pt} $\boxtimes$             Multi-Messenger Astronomy and Astrophysics \hspace*{65pt} \linebreak
  
\textbf{Principal Author:}

Name: Steven M. Kahn \linebreak						
Institution:  Stanford University, SLAC, LSST
 \linebreak
Email: SKahn@lsst.org \linebreak
 
\textbf{Co-authors:} (names and institutions)
  \linebreak
Robert Blum, AURA/LSST Operations\linebreak
Chuck Claver, AURA/LSST Project \linebreak  
Andrew Connolly, University of Washington \linebreak
Leanne Guy, AURA/LSST Project \linebreak  
\v{Z}eljko Ivezi\'{c}, University of Washington \linebreak  
Robert H. Lupton, Princeton University \linebreak
William O'Mullane, AURA/LSST Project \linebreak  
Steven Ritz, UC Santa Cruz and SCIPP \linebreak
Michael Strauss, Princeton University \linebreak
Christopher Stubbs, Harvard University \linebreak
Sandrine J. Thomas, AURA/LSST Project \linebreak
J. Anthony Tyson, UC Davis \linebreak
Bo Xin, AURA/LSST Project \linebreak  

\textbf{Abstract}
In this white paper, we discuss future uses of the LSST facility after the planned 10-year survey is complete. We expect the LSST survey to profoundly affect the scientific landscape over the next ten years, and it is likely that unexpected discoveries may drive its future scientific program. We discuss various operations and instrument options that could be considered for an extended LSST mission beyond ten years.
}

\pagebreak

\section{Introduction}
%\leftline{\Large 3 pages: Kahn}

The Large Synoptic Survey Telescope (LSST) will be a large-aperture, wide-field imaging survey of the entire southern hemisphere in six optical color bands (u, g, r, i, z, y). Over ten years of operations, LSST will make more than 825 visits to every part of the southern sky, where each visit will involve a single exposure or pair of exposures, roughly 30 s in duration.  Single visit limiting magnitudes range from 23.9 in the u-band to 22.1 in the y-band, yielding roughly 3 magnitudes greater depth for coadded images over the full ten years.  The science enabled by this survey will be very broad, ranging from studies of small moving bodies in the solar system to the structure and evolution of the universe as a whole. For a more detailed discussion of the science opportunities enabled by LSST, see \citealt[][]{2019ApJ...873..111I} (hereafter I19).

The nominal duration of the LSST survey will be ten years of continuous operation.  The science goals and requirements for the technical design assumed this duration, and the construction project is currently on track to achieve them.  However, given the very significant investment in LSST construction ($\sim$ \$700M), it makes sense to consider the fate of this world-unique facility after that ten-year nominal program has been completed.   The purpose of this white paper is to address that question from the viewpoint of the LSST Project Science Team, which represents the scientific leadership of the construction project.  Our goal is not to advocate for any particular scientific program, but merely to convey what we believe are the primary technical considerations that should be taken into account, as future potential uses are evaluated.  We are aware that there are other white papers that have been or will be submitted to Astro2020 arguing for particular choices based on various scientific motivations.  We have not been asked to evaluate those submissions, nor have we attempted to review them.

\section {Brief technical overview of the LSST design}

The large LSST \'etendue is achieved in a novel three-mirror design with a very fast $f$/1.234 beam. The optical 
design has been optimized to yield a large field of view (9.6 deg$^2$), 
with seeing-limited image quality, across a wide wavelength band (320--1050
nm). Incident light is collected by an annular primary mirror, having
an outer diameter of 8.4 m and inner diameter of 5.0 m, creating an effective filled aperture of 
$\sim$6.5 m in diameter. The collected light is reflected to a 3.4 m convex secondary, then onto
a 5 m concave tertiary, and finally  into the three refractive lenses of the camera.
%(see Fig.~\ref{Fig:optics}).  
In broad terms, the primary-secondary mirror pair acts as a beam condenser, while the aspheric portion of 
the secondary and tertiary mirror acts as a Schmidt camera.  The 3-element refractive optics of the camera
correct for the chromatic aberrations induced by the necessity of a thick dewar window and flatten the
focal surface.  All 
three mirrors are actively supported to control wavefront distortions 
introduced by gravity and environmental stresses on the telescope. 

The LSST Observing Facility,
consisting of the telescope enclosure and summit support building, is situated atop Cerro Pach\'{o}n in northern Chile,
sharing the ridge with the Gemini South and SOAR telescopes.  The telescope enclosure houses a compact, stiff
telescope structure atop a 15 m high concrete pier
with a fundamental frequency of 8 Hz, that is crucial for achieving the required fast slew-and-settle times.  The summit support building includes a coating chamber for recoating the three LSST mirrors and 
clean room facilities for maintaining and servicing the camera.

The LSST camera provides a 3.2 Gigapixel flat focal plane array, tiled by 189
4K$\times$4K CCD science sensors with 10 $\mu$m pixels. 
The sensors are deep depleted high resistivity silicon back-illuminated devices with 
a highly segmented architecture that enables the entire array to be read in 2 seconds. 
The detectors are grouped into 3$\times$3 rafts 
each of which 
contains its own dedicated electronics. The rafts are mounted on a silicon carbide 
grid inside a vacuum cryostat, with an intricate thermal control system that maintains 
the CCDs at an operating temperature of 173 K. The entrance window to the
cryostat is the third (L3) of the three refractive lenses in the camera. The other
two lenses (L1 and L2) are mounted in an optics structure at the front of the camera 
body, which also contains a mechanical shutter, and a carousel assembly that holds 
five large optical filters. The sixth optical filter can 
replace any of the five via a procedure accomplished during daylight hours. 

\section{Extended running with existing hardware}
%\leftline{\Large  4 pages: Ivezic, Blum, Strauss}

\subsection{Approximate operations costs at the end of the nominal LSST Survey} 
By 2032, LSST will be optimally run at its minimum level of effort. Future plans for extending or adding to the LSST Survey which are consistent with nominal operations will have a well understood cost and operations footprint. Adding new filters, or modifying the cadence or exposure times, for example would be straightforward to incorporate in the existing operations model. Such plans would have predictable impacts (greater or lesser) on the data management sizing model (more/less frequent images, for example lead to more/less storage and processing costs).

At the end of the main survey in 2032, LSST will have robust operations in Chile, Tucson, AZ, and at data centers in Illinois (NCSA), France, and perhaps elsewhere including in the cloud. The overall staffing level will be approximately 150 FTE over all locations employed by AURA and SLAC. 

Payroll and non-payroll costs are escalated in the operations plan (roughly 3\% per year), so the total cost is nearly constant in the out years of the survey as actual effort slowly ramps down until the end of 2032. The projected total operations cost is $\$$68M USD in then-year dollars for 2032. About $\$$44M of the total is in staff cost.  The distribution of costs to DOE and NSF in this model is the subject of current discussion at the agency level.

\subsection{Scaling with time of various LSST science cases for the WFD program}
\label{sec:duration}

We discuss here how data properties with direct impact on science deliverables 
vary with survey length, $t$.
The coadded depth (the 5$\sigma$ depth for point sources), $m_5^{\rm coadd}$, scales 
with time as (see eq.~6 in I19).  
\begin{equation} 
   m_5^{\rm coadd}  = 
   m_5^{\rm coadd, Final}  + 1.25 \, \log_{10}\left({t \over 10 \, {\rm yrs}}\right) 
\end{equation} 
where $m_5^{\rm coadd, Final}$ is the target depth achieved with 10-year survey (with
airmass losses taken into account, $m_5^{\rm coadd, Final}=27.5$ for the $r$ band). 

The photometric errors (inverse signal-to-noise ratio) at the faint limit of the so-called 
``gold'' galaxy sample (4 billion galaxies with $i<25$ which will be used for cosmological
programs), is computed from (see eq.~5 and Table 1 in I19):
\begin{equation} 
 \sigma_{i=25} = 0.04 \, \left({t \over 10 \, {\rm yrs}}\right)^{-1/2} {\rm mag.}
\end{equation}

The volume of the 5-dimensional color space per source with $i=25$,  which controls the 
ability to classify sources using colors (including photometric redshift estimates for 
galaxies and star/quasar separation, for example) is computed assuming uncorrelated 
color errors, as proportional to $\sigma^5_{i=25}$, and normalized by the value 
corresponding to the 10-year survey. 
The number of visits is assumed proportional to time, with a value of 825
corresponding to the main deep-wide-fast 10-year survey  (based on 
realistic cadence simulations that incorporate various losses and Cerro Pach\'on 
weather patterns). 

The trigonometric parallax accuracy for a point source with $r$=24 (see section 
3.2.3 in I19) scales with time as 
\begin{equation}
 \sigma_\pi = 3.0 \,  \left({t \over 10 \, {\rm yrs}}\right)^{-1/2}  \,\,  {\rm mas.} 
\end{equation}

The proper motion accuracy for a point source with $r$=24 (see section 3.3.3 in I19)
scales with time as 
\begin{equation}
 \sigma_\mu = 1.0 \,  \left({t \over 10 \, {\rm yrs}}\right)^{-3/2}   \,\, {\rm mas/yr.} 
\end{equation}
The very strong dependence of  $\sigma_\mu$ on time ($t^{-3/2}$) comes from the
increase in the square root of the number of visits, analogously to $\sigma_\pi$, 
and an additional $t^{-1}$ from the linear increase in temporal baseline.  

The behavior of these quantities as a function of time is summarized in 
Table~\ref{table:time}. As the table demonstrates, several important quantities 
show marked improvement between 
the survey years 8 and 10: most notably, the color volume per source for faint sources 
($i=25$) shrinks by a factor of 1.7. Substantial improvement is also seen for proper 
motions, with errors larger by 40\%, after 8 years than at the end of the 10-year survey. 
These quantities would benefit the most from continuing surveying beyond the 10-year limit.

An incomplete list of science programs with figures of merits that 
would continue improving with survey duration include the completeness
for hazardous asteroids \citep{2018Icar..303..181J}, completeness for periodic 
variable stars \citep{2015ApJ...812...18V}, and dark energy science, described below.

\begin{table}
\centering
\caption{Various science metrics as functions of survey duration. \label{table:time}}
\begin{tabular}{|l|r|r|r|r|r|r|}
\hline     
Quantity                          &     Year 1   &    Y3  &     Y5  &     Y8   &     Year 10   &     Y12  \\
\hline  
    $r_5$ coadd$^a$                   &       26.3    &      26.8   &      27.1    &      27.4    &          27.5   &       27.6    \\
    $\sigma$($i$=25)$^b$         &     0.12    &     0.07    &      0.06    &    0.05      &        0.04     &      0.04     \\     
    color vol.$^c$                        &       316     &       20     &        6      &    1.7        &           1       &        0.6    \\
     \# of visits$^d$                    &          83     &     248     &      412     &    660      &          825    &       990      \\  
    $\sigma_\pi$ ($r$=24)$^e$   &        9.5     &      5.5     &        4.2    &     3.3       &          3.0     &        2.7      \\ 
    $\sigma_\mu$ ($r$=24) $^f$  &  32   &      6.1    &     2.8   &     1.4   &     1.0   &       0.8     \\
\hline                         
\end{tabular}
\\ \vskip 0.05in

$^a$ The coadded depth in the $r$ band (AB, 5$\sigma$; point sources).  \\
$^b$ The photometric error for a point source with $i=25$. \\
$^c$ The volume of the 5-dimensional color space, normalized by the final value. \\
$^d$ The number of visits per sky position (summed over all bands). \\
$^e$ The trigonometric parallax accuracy for a point source with $r$=24 (milliarcsec). \\ 
$^f$  The proper motion accuracy for a point source with $r$=24 (milliarcsec/yr).  \\
%\vskip 0.2in          
\end{table}

\subsubsection{Dark Energy Figure of Merit as a Function of Survey Duration}

As a representative program that combines most aspects of LSST data, from coadded 
depth to time domain parameters, we consider the contribution of various observational 
probes into the nature of dark energy. We use the Dark Energy Task 
Force (DETF) figure of merit\footnote{DETF defined a dark-energy ``figure of merit''
using a two-parameter model for the dark energy equation of state, $w(a)$ = $w_0 + (1-a)\,w_a$, where $w_0$ is the present value of $w$, and where $w_a$ parameterizes the evolution of 
$w(a)$. The DETF FOM is the reciprocal of the area of the error ellipse enclosing the 95\% confidence 
limit in the $w_0$-$w_a$ plane. Larger figure of merit indicates greater accuracy. For more
details, please see the DETF Report.} (FOM) as a proxy for 
dark energy capability. The precision for cosmological parameters from LSST is a strong 
function of time, reflecting both decreasing random errors in galaxy photometry and shape 
measurements, as well as a reduction in various systematic errors due to a combination of
multiple cosmological probes (baryon acoustic oscillations, weak lensing correlations, and supernovae). We find that the DETF error product continues to decrease approximately linearly
with $t^{-1}$ as a function of time, even in the last two years of the survey. 

This is driven by weak lensing and BAO signal-to-noise ratios. The errors of the two 
quantities that enter the DETF FOM as a product
each decrease proportionally to $t^{-1/2}$. This error behavior is driven by random 
shear errors and assumes that systematic shear errors are reduced to an essentially 
negligible level of $\sim10^{-7}$, or about 3 times smaller than random shear errors
by the end of the survey (the required shear error level is set by the cosmological signal 
strength and the desired value of the DETF FOM; the systematic shear errors for a single 
visit, at the level of 10$^{-5}$, are reduced to their final negligible value due to 
instrument signature removal during processing and averaging
of at least 100 position and angle dithered visits per band).

\subsection{Special (non-canonical) observing campaigns}

With its enormous imaging \'etendue, the nominal LSST program enables a broad range of science with a single observing mode.  The so-called ``Wide-Fast-Deep'' mode is designed to repeatedly cover $\sim 20,000$ deg$^2$ in the six canonical bands over the ten-year survey.  While a default observing strategy and cadence will yield the raw data needed to carry out the primary scientific programs, it is clear that certain science cases would benefit significantly from alternative approaches to survey strategy. The current plan does allow for about 10\% of the time to be devoted to special programs. A recent call to the community for ideas of how to use that time resulted in 46 white papers covering all aspects of LSST science.  It is clear from the range of ideas proposed that even with no change in instrumentation, LSST in an extended mission could be used with fundamentally new observing strategies to achieve a large set of new and different science goals. 

 Most of the ideas submitted involve variations from the default cadence with which LSST covers the sky.  The current default WFD cadence allows repeated observations of a given area of sky roughly once every three nights.  The variable universe at LSST single-visit depths ($r\sim24.5$) on significantly shorter timescales than that is completely unexplored.  The choice of how the different filters would be deployed in a much more rapid cadence, over a more limited region of sky, needs to be explored, and will depend in part on the science goals, and whether one is observing at low or high Galactic latitude.  
  
  Another example is the follow-up of transient events discovered with other facilities, especially gravitational wave detectors. LSST is likely to be the premier facility for finding optical counterparts to LIGO-VIRGO-KAGRA sources for some time.  But the enormous progress in the field over the last few years suggests that the landscape will be very different a decade from now, and it is hard to predict what the most exciting opportunities will be for follow-up of gravitational wave sources.

\section{Modest cost modifications of the existing hardware}
%\leftline{\Large  3 pages:  Ritz, Connolly, Tyson, Ivezic}

\subsection{Maintenance and refurbishment of the LSST camera}
Extending the camera lifetime beyond the ten-year science survey and/or making modest improvements should be fairly straightforward. For lifetime extension, the first consideration would be components with moving parts, which are generally designed to last 15 years. An assessment of actual parts wear can be made during the survey in scheduled maintenance periods. This category includes the shutter, which is easily accessible; the various filter exchange mechanisms, some of which might require more extensive replacement after the survey is complete; pumps, compressors, and refrigerant; and the hexapod/rotator. Most of the other hardware, including the filter coatings, sensors, electronics, power supplies, and data acquisition and control systems are not expected to degrade substantially, but are all serviceable on various time scales. It is reasonable to expect some loss of channels over time. Since there are 3,024 CCD channels in the science focal plane, the loss of a modest number of  channels would not likely require immediate action. 

Replacing focal plane sensors is possible. The main focal plane array is modular, with 9 sensors and associated electronics on each of 21 mechanically separate science rafts, which mount to a flat grid. The LSST operations plan includes maintaining the capabilities needed to replace and service science rafts as well as the four corner rafts that support the guide sensors and wavefront sensors. Opening the cryostat and replacing several or more rafts would likely take at least two months, and there are associated risks, so the (re)gain of science capability would be carefully considered in context of those risks and the potential loss of observing time. Replacing the sensors with devices that have better IR sensitivity (see below), and associated electronics if needed, could be done for the full focal plane or for a subset, as individual rafts can be replaced. Implementation of different thermal requirements would potentially be more complicated. For a mixed focal plane, the installed $z$ height of the new sensor assembly would have to be carefully matched to that of the rest of the focal plane. 

Enhancing the filter complement (see below) is straightforward and low risk, as the design supports routine changing of one or more of the on-board filters as needed.

\subsection{Change-out of the LSST filter set}

Replacement of the LSST filter set is an obvious, relatively simple modification that could be made to the LSST camera, yielding new science potential for an extended observing campaign.  Given the cost of the original LSST filters, we estimate that such a replacement will be in the range \$5-10M, depending on how many new filters are produced, and whether the existing filter blanks are made available for recoating, or whether new blanks would have to be fabricated.  Below we consider two possibilities:  (1) fabrication of a set of complementary broadband filters; and (2) fabrication of a set of narrowband filters.

\subsubsection{New set of 1-5 bands}
\label{sec:broad}

Calibration of photometric redshifts, and other elements of SED estimation from multi-color measurements, is likely to remain a key element of LSST science, even after ten years.  One possible modification to the system could involve a new filter complement to aid in these determinations for the sources that have already been surveyed.   To design optimal filters for photometric redshifts it is useful to calculate the information gain to quantify how well a filter separates the colors measured for a set of templates at given redshift from the colors measured at other redshifts. 
A modified set of wideband filters -- shifted by up to half the single filter bandwidth -- would produce better SED information on faint galaxies at the same high S/N as the original set. The photo-z degeneracies produced by the existing five inter-filter transitions would be partially broken, resulting in a decreased outlier fraction.

If the detector  remains silicon, the optimal deployment of the 350-1000 nm high-QE region would be filters interspersed between the ugrizy bands. An example 5 filter set would be shifted filters centered on the 5 transition wavelengths: ug, gr, ri, iz, zy. That full depth survey would take 8 years.  However there are attractive alternatives ranging from one additional filter to several. %Figure~\ref{Fig:modfilt-pzoutliers} shows the change in photo-z outlier rate for the addition of one filter and for an alternate 6 filter set.

%\begin{figure}[]
%\vskip -0.8in
%\begin{center}
%\includegraphics[width=0.6\hsize,clip]{outlier_fraction.png}
%\vskip -0.2in
%\caption{The photo-z outlier fraction for three modified filter scenarios: the original LSST filters, the LSST filters with an additional single broadband filter, a new set of six LSST filters} 
%\label{Fig:modfilt-pzoutliers}
%\end{center}

%\end{figure}

%An example of the impact of new filters can be seen in Figure~\ref{Fig:modfilt-pzoutliers}, which shows the change in photo-z outlier rate. For one extra filter, Bryce Kalmbach normalized the sources to $i$=25 and set a prior on redshift with a peak of around 0.92 to match the galaxies expected at this magnitude. Locations of the four corners of a trapezoidal filter were allowed to vary independently over the wavelength range 300 - 1100 nm. This optimizes the shape and location of the new filter. At each optimization step the information gain was calculated using the colors from the current LSST filters along with the new filter. The resulting  optimized filter is quite broad:  the best 7th filter for photo-z focuses on the location of the 4000 angstrom break at the peak of the redshift prior and is set between the $r$ and $i$ LSST filters. 

\subsubsection{Narrow band filter sets} 

There are obvious advantages to potentially complementing the main survey with a set of narrowband measurements
\citep{2019arXiv190306834Y}. The fast f/1.2 beam of LSST makes it complicated but not impossible to incorporate traditional narrowband interference filters (with the bandpass width limited to about 10-20 nm). 

The LSST beam at the filter location is a hollow cone of rays, with the (virtual) chief ray normal to the filter surface.  The filter is 150 mm from the focal plane, on axis. The $f$/1.23 annular beam footprint therefore has an outer diameter at the filter that spans about 18 cm. There is a 1:1 correlation between the ray angle and its radial position relative to the (virtual) chief ray at the center of the beam. The angle of incidence on the filter ranges from 14 to 23 degrees relative to the normal to the filter surface, with the 23 degree rays striking the filter at the outer edge of the annular beam footprint. 

Thin-film interference filters suffer an angle-of-incidence dependent shift in transmission properties, with passband edges shifting to bluer wavelengths by an amount approximated by 
\begin{equation}
   \lambda(\theta)=\lambda_0 \sqrt{1-(\sin\theta/n_{eff})^2}, 
\end{equation}
where $n_{eff}$ is the effective index of refraction of the thin film layers and $\theta$ is the angle off normal. Typical values of $n_{eff}$ are in the range of 1.5 to 2.5.   

For $n_{eff}$ = 1.8, the incidence-angle-dependent shift in wavelength change ranges from 0.976 to 0.991 (compared to normal incidence), a span of 1.5\%. Note that these rays are not equally weighted. There are more photons impinging at 23 degrees than at 14 degrees. So a delta-function normal incidence filter produces a skewed, blue-shifted response in the LSST beam. For the longest wavelength narrowband filter we might imagine placing in the beam, centered at 1000 nm, this limits our bandwidth to 15 nm, convolved with the normal-incidence response function of the filter. At 500 nm the angle-driven convolution width is half this value, about 7.5 nm. If we fabricate filters with higher effective indices, the effect is attenuated. At neff=2.2 the fractional broadening is (0.944  --0.984)=1\%, which equates to 5 nm at our midpoint wavelength of 500 nm.  

Given the size and shape of the LSST filters, it is probably impractical to imagine producing a filter with a uniform normal-incidence width of under 10 nm. With a high-index thin film narrowband interference filter, we can probably expect to achieve a transmission FHWM of 15-20 nm in the LSST beam. This is about a factor of 7 narrower than the typical LSST broad passband.

\section{Replacement of the camera with an alternate instrument}
%\leftline{\Large  5 pages:  Stubbs, Thomas}

\subsection{Multiobject spectrograph}

The \'etendue of the LSST optical system is potentially attractive for spectroscopy, but only if the full field is populated with optical fibers at high packing density. For the comparison of spectroscopic merit, the imaging \'etendue A$\Omega$ is replaced by AN$_{fibers}$, where N$_{fibers}$ is the total number of optical fibers that populate the focal plane. 

The spatial density of objects with $20<i_{AB}<23.5$ is 14 per square arcminute, or $\sim$450,000 per LSST field of view. One conceptual approach would be to invest the resources needed to capture all of these in a single pointing, with a half-million object spectrograph. That would require a tremendous investment in the instrument, with a short data collection period.  Taking a rough estimate of \$50M/yr for LST operating costs and an instrument investment of \$500M implies a ten year spectroscopic survey. This means that each field can be revisited multiple times, adopting 10 pointings as a goal. One could then have ten fiber configurations with which to observe the targets of interest, which in turn implies about 40k optical fibers in the focal plane. 

Taking a target spectral resolution of R=$\lambda/\Delta\lambda \sim$ 10,000, with 7 pixels between spectra in the spatial direction, with good sampling in the spectral direction a 30,000 object spectrograph would require a total number of pixels $N_{pix}=2\times10,000\times30K\times7\times NF$, where $NF$ is the number of fibers allocated per object.  An $NF \sim 1.2$ (as assumed by \textit{e.g.} PFS, \citealt{PFSSpectrograph}) would probably provide suppression of systematic errors that arise from sky subtraction and atmospheric dispersion, giving $N_{pix}\sim$ 12 Gpix; if $NF$ is in reality $3-5$, the number of pixels would be a few times larger.

The comparatively short focal length of LSST provides a plate scale that is more compact than is typical on most other large-aperture telescopes. That fact, in conjunction with the strategy of having $\sim$ 10 distinct pointings per field, means that the full-range-of-motion required for an LSST fiber positioner is of order 1mm, about a factor of 5-10 smaller than current practice. As the objects of interest for spectroscopy push to fainter levels, the tolerable fractional error in sky subtraction becomes more demanding. 
%This is a topic for potential R\&D efforts, since none of the existing fiber positioners operate in this regime. \XXX{I think you'd want to be able to reach any point, so the patrol region diameter is 60cm/$\sqrt{40k} \sim 3-4mm$ minimum, within a factor of 3 of PFS}
%Two potential sources of systematic error were discussed: differential atmospheric refraction and sky subtraction artifacts. 

%The determination of the appropriate sky spectrum for each object is a combined hardware+software challenge. Subtle details about the pattern of illumination on the optical fiber tip, and whether the fiber is best illuminated in the image plane or with a re-imaged pupil were topics of discussion. There was considerable sympathy for the notion of allocating fibers to each object for local sky determination, perhaps as a 1-d array. \XXX{I am not convinced that this is necessary.  PFS will find out, but the AAT folk are claiming scaling as sqrt(N) out to 40 hour integrations}
Another challenge is the lack of an atmospheric dispersion corrector. Although the distortion of the spectrum due to differential chromatic refraction (DCR) can be ameliorated somewhat using the broadband imaging data, for fiber diameters smaller than the refracted footprint the photons are lost and the signal-to-noise ratio suffers. Making the fiber large enough to capture the entire DCR-displaced footprint introduces unwanted shot noise from the sky. 
%This implies that a fiber-by-fiber correction for atmospheric dispersion would be very beneficial. 

\subsection{Infrared camera}

The blue end of the current LSST wavelength range is set by the cutoff in optical transmission of the atmosphere, but the red end is set by the bandgap of the silicon CCDs. Changing to an alternative detector technology could increase the wavelength coverage in the red. There is little gain in extending the wavelength range beyond 1.8 or 2.0 microns, since the LSST has no cold pupil stop and the thermal background rate will degrade the achievable signal-to-noise ratio. So we adopt 1.8 microns as a good goal for an upper bound on the near infrared wavelength. Using the current generation of HgCdTe devices (Hawaii 4 RGs) to pave the focal plane would be very expensive. 

CCDs made of Germanium would extend into this range, and there are at least two development efforts under way to construct Ge CCD sensors. This is a potentially viable path, if the sensor costs can be brought down to a few pennies per pixel. Again, unless the full etendue of the LSST optical design were exploited it makes more sense to populate the focal plane of a small field of view telescope with this emerging technology. 

There is one infrared sensor material, InGaAs, which is well-suited to imaging out to 1.7 microns. The WINTER project for instance is filling a 1x1 degree field with InGaAs which is 10\% of the LSST field, and will serve as a valuable precursor project. The main challenges to implementing an InGaAs focal plane for LSST are developing astronomy-grade sensors that can be readily assembled into a high-fill-factor configuration, with acceptably low read noise and dark current. The state of the art of such sensors is promising, with the prospect of operating this focal plane at a temperature comparable to that of the current LSST focal plane. This suggests that the thermal management problem might be tractable. The LSST readout electronics would have to be replaced with a highly parallelized system appropriate for multiplexed pixel architectures. 

In addition, the coating of the lenses is optimized for wavelengths from 350nm to 1.1 microns, with a sharp cut off after 1 micron. This implies that the coating would have to be redone for wavelenghts up to 1.8 microns. Considering the complexity of removing the lenses from the camera body and the time/risk it represents to recoat the lenses, it would be safer to redo the lenses in advance of the swap. 
Considering the large set of changes required, and the uncertain state of this technology, the costs of such an instrument replacements are likely to be considerable, at least several hundred million dollars.

\subsection{sCMOS camera}

A key potential science opportunity for an extended LSST survey would be to enhance the discovery potential for very short-timescale phenomena. Several aspects of the existing LSST CCD camera prevent fast exposures.  The primary constraint is mechanical: a large mechanical shutter is required for the frame transfer read. The current LSST camera electronics and data processing are designed around frame transfer CCDs. This is inherently a relatively slow process: even with parallel reads of the 16 segments of the LSST 16 Mpixel CCDs, the read noise rises for total read times less than 2 seconds, making exposures less than 10-20 seconds inefficient. CMOS cameras on the other hand can be designed differently, with fast electronics for motion detection, for example. Applied to a 2nd generation LSST camera, the CMOS electronics and digital logic would reside next to each CMOS array, with only optical fibers exiting the dewar.  The power requirements would be much smaller than the existing LSST camera.

While 3-4 micron pixel CMOS imagers have been developed for consumer applications, a new family of low noise high-QE scientific CMOS detectors, sCMOS, has recently been undergoing rapid development as well.  The singular advantages of this type of detector are sub-second low noise read, and self shuttering -- so that the camera would not have to incorporate a faster shutter. These sensors feature kHz frame rates and support non-destructive read, enabling lower read noise on second timescales. 

Scientific CMOS development is accelerating. Medical imaging applications are driving large format, and quantum computing applications are driving high QE and rapid read.  There has been recent progress in back-illuminated sCMOS.
%Andor has 11 micron pixel back-illuminated sCMOS with 95\% QE now measuring 3 cm, and the market is for much larger detectors.  
Whether via mosaicing or larger individual sensors, it is likely that much larger sCMOS will be developed by 2030. Embedded signal processing is becoming more common, and by the end of the LSST survey this new class of intelligent imager could emerge as an attractive choice for a follow-on mission for LSST.
The costs of such a replacement are too uncertain at this stage to estimate, in light of the rapid state of technological development.  However, it is likely that this would require an investment at least $\sim$\$100M.

\section{Summary and conclusions}

We have discussed various options that could be considered for an extended LSST mission beyond ten years, some of which might involve only modest construction costs, and some of which are much more significant.  In all cases, there will be continued operational costs that would remain as a major component of the NSF and DOE annual budgets.  If international partners can be recruited, some of the operational costs could potentially be mitigated, although at the expense of added complication.

It should be emphasized that decommissioning LSST is also not without significant cost.  Simply mothballing the facility could present safety hazards, so this would have to be carefully studied.  We do not presently have estimates of what it would take to dismantle the facility and return the site to an alternate state, but we expect those costs to be very considerable.

While the Astro2020 deliberations will have to be conducted over the next year or so, we also recognize that the full scientific benefits of any of these plans will be much better understood once LSST is operating, and we have a more complete understanding of its performance and programmatic characteristics.  In addition, we are expecting the nominal LSST program to profoundly affect the scientific landscape over the next ten years, and it is likely that the unexpected discoveries that it will make may drive its future scientific program.  So this is an uncertain process, but we hope that the comments above are useful for the Committee's considerations.
 
This work was performed in part at Aspen Center for Physics, which is supported by National Science Foundation grant PHY-1607611.

%\clearpage
\let\oldbibliography\thebibliography
\renewcommand{\thebibliography}[1]{\oldbibliography{#1}
\setlength{\itemsep}{0pt}} %Reducing spacing in the bibliography.
\bibliography{ref}

\begin{thebibliography}{}
\expandafter\ifx\csname natexlab\endcsname\relax\def\natexlab#1{#1}\fi

\bibitem[{{Ivezi{\'c}} {et~al.}(2019){Ivezi{\'c}}, {Kahn}, {Tyson}, {Abel},
  {Acosta}, {Allsman}, {Alonso}, {AlSayyad}, {Anderson}, {Andrew}, \&
  et~al.}]{2019ApJ...873..111I}
{Ivezi{\'c}}, {\v Z}., {Kahn}, S.~M., {Tyson}, J.~A., {et~al.} 2019, \apj, 873,
  111

\bibitem[{{Jones} {et~al.}(2018){Jones}, {Slater}, {Moeyens}, {Allen},
  {Axelrod}, {Cook}, {Ivezi{\'c}}, {Juri{\'c}}, {Myers}, \&
  {Petry}}]{2018Icar..303..181J}
{Jones}, R.~L., {Slater}, C.~T., {Moeyens}, J., {et~al.} 2018, Icarus, 303, 181

\bibitem[{{Tamura} {et~al.}(2018){Tamura}, {Takato}, {Shimono}, {Moritani},
  {Yabe}, {Ishizuka}, {Kamata}, {Ueda}, {Aghazarian}, {Arnouts}, {Barkhouser},
  {Balard}, {Barette}, {Belhadi}, {Burnham}, {Caplar}, {Carr}, {Chabaud},
  {Chang}, {Chen}, {Chou}, {Chu}, {Cohen}, {de Almeida}, {de Oliveira}, {de
  Oliveira}, {Dekany}, {Dohlen}, {dos Santos}, {dos Santos}, {Ellis},
  {Fabricius}, {Ferreira}, {Furusawa}, {Garcia-Carpio}, {Golebiowski}, {Gross},
  {Gunn}, {Hammond}, {Harding}, {Hart}, {Heckman}, {Ho}, {Hope}, {Hover},
  {Hsu}, {Hu}, {Huang}, {Jamal}, {Jaquet}, {Jeschke}, {Jing}, {Kado-Fong},
  {Karr}, {Kimura}, {King}, {Koike}, {Komatsu}, {Le Brun}, {Le F{\`e}vre}, {Le
  Fur}, {Le Mignant}, {Ling}, {Loomis}, {Lupton}, {Madec}, {Mao}, {Marchesini},
  {Marrara}, {Medvedev}, {Mineo}, {Minowa}, {Murayama}, {Murray}, {Ohyama},
  {Onodera}, {Orndorff}, {Pascal}, {Peebles}, {Pernot}, {Pourcelot}, {Reiley},
  {Reinecke}, {Roberts}, {Rosa}, {Rousselle}, {Schmitt}, {Schwochert},
  {Seiffert}, {Siddiqui}, {Smee}, {Sodr{\'e}}, {Steinkraus}, {Strauss},
  {Surace}, {Tait}, {Takada}, {Tamura}, {Tanaka}, {Tanaka}, {Thakar},
  {Verducci}, {Vibert}, {Wang}, {Wang}, {Wen}, {Werner}, {Yamada}, {Yan},
  {Yasuda}, {Yoshida}, \& {Yoshida}}]{PFSSpectrograph}
{Tamura}, N., {Takato}, N., {Shimono}, A., {et~al.} 2018, in Society of
  Photo-Optical Instrumentation Engineers (SPIE) Conference Series, Vol. 10702,
  107021C

\bibitem[{{VanderPlas} \& {Ivezi{\'c}}(2015)}]{2015ApJ...812...18V}
{VanderPlas}, J.~T., \& {Ivezi{\'c}}, {\v Z}. 2015, \apj, 812, 18

\bibitem[{{Yoachim} {et~al.}(2019){Yoachim}, {Graham}, {Bet}, {Vu{\v
  c}eti{\'c}}, {Ivezi{\'c}}, {Boyer}, {Arbutina}, \&
  {Jones}}]{2019arXiv190306834Y}
{Yoachim}, P., {Graham}, M., {Bet}, S., {et~al.} 2019, arXiv e-prints,
  arXiv:1903.06834

\end{thebibliography}

\end{document}